\documentclass{iaus}
\usepackage{graphicx}

\newcommand{\be}{\begin{equation}}
\newcommand{\ee}{\end{equation}}
\newcommand{\beqa}{\begin{eqnarray}}
\newcommand{\eeqa}{\end{eqnarray}}

\def\Mo{{\rm M_\odot}}

\def\kpc{\ {\rm kpc}}
\def\pc{\ {\rm pc}}
\def\kms{{\ }{\rm km}\,{\rm s}^{-1}}
\def\LCDM{$\Lambda$CDM}
\def\degrees{^\circ}

\title[CDM Substructure and Galactic Disks] 
{Cold Dark Matter Substructure \\ and Galactic Disks}

\author[Kazantzidis et al.]   
{Stelios Kazantzidis$^1$, Andrew R. Zentner$^2$, \and James S. Bullock$^3$}

\affiliation{$^1$Center for Cosmology and Astro-Particle Physics, The Ohio
  State University, \\ 191 West Woodruff Avenue, Columbus, OH 43210, USA 
\\email: {\tt stelios@mps.ohio-state.edu} \\[\affilskip]
$^2$Department of Physics \& Astronomy, University of Pittsburgh, \\
100 Allen Hall, 3941 O'Hara Street, Pittsburgh, PA 15260, USA 
\\email: {\tt zentner@pitt.edu} \\[\affilskip]
$^3$Center for Cosmology, Department of Physics \& Astronomy, \\ 
The University of California at Irvine, 4168 Reines Hall, Irvine, CA 92697, USA 
\\email: {\tt bullock@uci.edu}}

\pubyear{2008}
\volume{254}  
\pagerange{1--6}
\jname{The Galaxy Disk in Cosmological Context}
\editors{J. Andersen, J. Bland-Hawthorn \& B. Nordstr\"om, eds.}
\begin{document}

\maketitle

\begin{abstract}

We perform a set of high-resolution, fully self-consistent dissipationless $N$-body 
simulations to investigate the influence of cold dark matter (CDM)
substructure on the dynamical evolution of thin galactic disks. Our 
method combines cosmological simulations of galaxy-sized CDM 
halos to derive the properties of substructure populations and
controlled numerical experiments of consecutive subhalo impacts onto
initially-thin, fully-formed disk galaxies. We demonstrate that close
encounters between massive subhalos and galactic disks since $z \sim 1$
should be common occurrences in {\LCDM} models. In contrast,
extremely few satellites in present-day CDM halos are likely to have a
significant impact on the disk structure. One typical host halo merger 
history is used to seed controlled $N$-body experiments of subhalo-disk 
encounters. As a result of these accretion events, the
disk thickens considerably at all radii with the disk scale height
increasing in excess of a factor of $2$ in the solar neighborhood. We 
show that interactions with the subhalo population produce a wealth of distinctive morphological 
signatures in the disk stars, many of which resemble those being discovered 
in the Milky Way (MW), M31, and in other disk galaxies, including: conspicuous flares; bars;
low-lived, ring-like features in the outskirts; and low-density, filamentary structures 
above the disk plane. We compare a resulting dynamically-cold, ring-like feature
in our simulations to the Monoceros ring stellar structure in the MW. The comparison shows 
quantitative agreement in both spatial distribution and kinematics, suggesting that such 
observed complex stellar components may arise naturally as disk stars are excited by 
encounters with subhalos.  These findings highlight the
significant role of CDM substructure in setting the structure of disk
galaxies and driving galaxy evolution.

\keywords{cosmology: theory, dark matter, galaxies: formation, galaxies:
  dynamics, galaxies: structure, methods: numerical}
\end{abstract}

\firstsection 
              
\section{Introduction}

The currently favored cold dark matter (CDM) paradigm of hierarchical 
structure formation (e.g., \cite[Blumenthal et al. 1984]{Blumenthal_etal84}), 
predicts significant dark matter halo substructure in the form of small, 
dense, self-bound {\it subhalos} orbiting within the virialized
regions of larger host halos (e.g., \cite[Klypin et al. 1999]{Klypin_etal99}). 
Observational probes of substructure abundance thus constitute fundamental 
tests of the CDM model. Due to the fact that most subhalos associated with galaxy-sized 
host halos lack of a significant luminous component, a constraint on the
amount of substructure in these systems may be obtained via their
gravitational influence on galactic disks. If there is a considerable subhalo 
population, it may produce strong tidal effects and induce distinctive gravitational signatures 
which might be imprinted on the structure and kinematics of the host galactic disk. 
Thus, establishing the role of substructure in shaping the fine 
structure of galactic disks may prove fundamental in informing our ideas 
about global properties of galaxy formation and evolution. 

Significant theoretical effort has been devoted 
to quantifying the resilience of galactic disks to infalling satellites
(e.g., \cite[Quinn \& Goodman 1986; Velazquez \& White
1999; Font et al. 2001; Gauthier et al. 2006; Read et al. 2008; 
Villalobos \& Helmi 2008]{Quinn_Goodman86,Walker_etal96,Velazquez_White99,Font_etal01,
Gauthier_etal06,Read_etal08,Villalobos_Helmi08}). Despite their usefulness, 
most earlier investigations suffered basic shortcomings that limited their 
applicability. For example, some considered encounters of single satellites 
with galactic disks, a set-up which is at odds with CDM predictions 
of multiple, nearly contemporaneous accretion events. Other studies made ad hoc assumptions 
about the orbital parameters and internal structures of the infalling systems. 
Consequently, it remains uncertain whether these earlier investigations faithfully 
captured the responses of galactic disks to halo substructure in a cosmological context.

Here we address this issue using a hybrid approach that 
combines cosmological simulations to derive the merger
histories of galaxy-sized CDM halos with controlled numerical experiments 
of consecutive subhalo impacts onto $N$-body realizations of fully-formed 
disk galaxies. We demonstrate that accretion histories of the kind 
expected in {\LCDM} models are capable of severely perturbing galactic disks 
and generating a wealth of distinctive signatures in the structural and kinematic 
properties of disk stars. The resulting morphological features are similar to those being
discovered in the Milky Way (MW), M31, and in other disk galaxies. 
We also show that satellite-disk interactions produce dynamically-cold,
ring-like features around galactic disks that are quantitatively 
similar to the Monoceros ring in the MW. This suggests that such 
observed stellar structures may arise naturally as a result of 
subhalo-disk encounters, which can excite perturbations in {\it disk stars}.
These findings imply that detailed observations of galactic structure may be able to distinguish between
competing cosmological models by determining whether the detailed structure 
of galactic disks is as excited as predicted by the CDM paradigm. 

\section{Methods}

The aim of this study is to assess the effects of CDM substructure 
on the dynamical evolution of thin galactic disks. A thorough 
description of our methods is presented in \cite[Kazantzidis et al. (2007)]
{Kazantzidis_etal07} and we summarize them here. 
First, we analyze cosmological simulations of the formation of four galaxy-sized 
halos in the {\LCDM} cosmology. The simulations were performed with the 
Adaptive Refinement Tree (ART) $N$-body code (\cite[Kravtsov 1999]{Kravtsov99}).
All of these halos accrete only a small fraction of their final 
masses and experience no major mergers at $z \lesssim 1$ (a look-back time of
$\approx 8$~Gyr), and therefore may reasonably host a disk galaxy. Second, 
we identify subhalos in these hosts and select the massive substructures 
that pass near the center of the host halo where they may interact appreciably 
with a galactic disk for further consideration. Finally, we use a representative 
subset of these accretion events from one of the host halos to seed controlled
$N$-body simulations of satellite impacts onto an initially-thin disk galaxy.

While the present work is informed by many past numerical investigations of
satellite-disk interactions, our methodology is characterized by at least
three major improvements. First, we consider satellite populations
whose properties (mass functions, internal structures, orbital parameters, 
and accretion times) are extracted directly from the cosmological 
simulations of galaxy-sized CDM halos. This eliminates many assumptions 
regarding the internal properties and impact parameters of infalling 
systems inherent in many previous studies. Second, we employ
primary disk galaxy models that are both self-consistent and flexible enough 
to permit detailed modeling of actual galaxies such as the MW and M31 by fitting to a wide 
range of observational data sets (\cite[Widrow \& Dubinski 2005]{Widrow_Dubinski05}). 
This allows us to set up thin, equilibrium realistic disk galaxies without 
instabilities and transient effects that can be present in other schemes. 
In this work, we employ galaxy model MWb of \cite[Widrow \& Dubinski (2005)]
{Widrow_Dubinski05} which reproduces many of the observed characteristics 
of the MW galaxy. This galaxy model comprises an exponential stellar disk  
with a sech$^2$ scale height of $z_d = 400\pc$, a Hernquist model bulge, 
and an NFW dark matter halo.

Lastly, and most importantly, we incorporate for the first time a model in
which the infalling subhalo populations are representative of those
that impinge upon halo centers since $z \sim 1$, instead of the $z=0$ 
{\it surviving} substructure present in a CDM halo. Previous studies 
utilized the {\it present-day} properties of a large ensemble of dark matter 
subhalos in order to investigate the dynamical effects of substructure on
galactic disks (\cite[Font et al. 2001; Gauthier et al. 2006]{Font_etal01,Gauthier_etal06}).
Successes notwithstanding, this methodology has the drawback 
of eliminating from consideration those massive satellites that, prior to $z=0$, pass 
very close to the central regions of their hosts, where the galactic disk resides.
These systems can potentially produce strong tidal effects on the disk, but
are unlikely to constitute effective perturbers at $z=0$ as 
they suffer substantial mass loss during their orbital evolution precisely 
because of their forays into the central halo 
(e.g., \cite[Zentner \& Bullock 2003]{Zentner_Bullock03}). 

\begin{figure}
\begin{center}
 \includegraphics[width=4in]{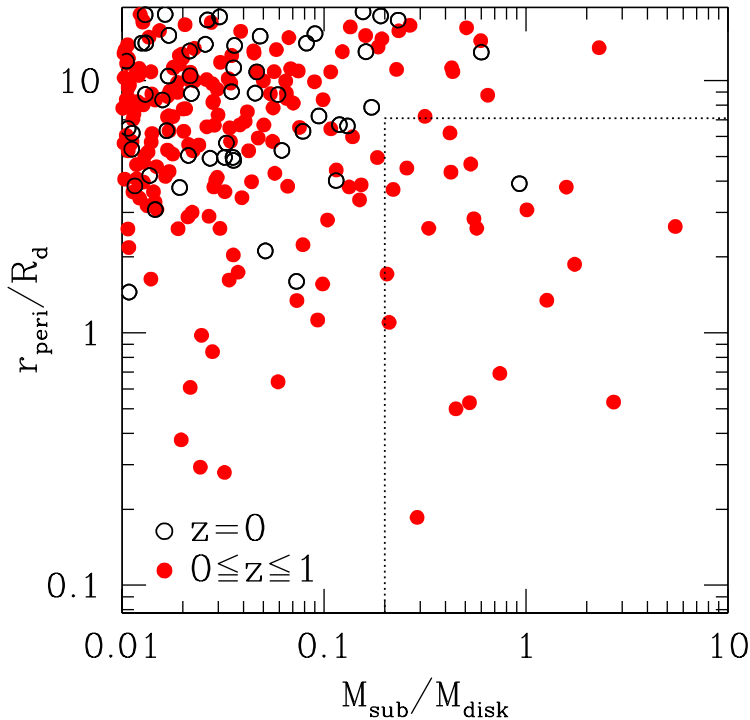} 
 \caption{A scatter plot of mass versus pericentric distance for satellites
  identified in four galaxy-sized halos formed in the {\LCDM} cosmology. Subhalo 
  masses and orbital radii are presented in units of the mass, 
  $M_{\rm disk}=3.53 \times 10^{10} \Mo$, and radial scale length, $R_d = 2.82\kpc$, 
  respectively, of the galactic disk in the controlled simulations. 
  {\it Filled} symbols show results for subhalos that pass closer than an 
  infall radius of $r_{\rm inf}=50\kpc$ of their host halo center since 
  a redshift $z = 1$. {\it Open} symbols refer to the $z=0$ population of
  surviving substructures. {\it Dotted} lines mark the so-called ``danger
  zone'' ($M_{\rm sub} \gtrsim 0.2 M_{\rm disk}$ and $r_{\rm peri} \lesssim
  20\kpc$) corresponding to infalling subhalos that are likely 
  capable of substantially perturbing the galactic disk. Accretions of massive
  subhalos onto the central regions of their hosts, where the galactic
  disk resides, since $z \sim 1$ should be common occurrences in standard
  {\LCDM}. In contrast, very few satellites in present-day subhalo populations
  are likely to have a significant dynamical impact on the disk structure.}
   \label{fig1}
\end{center}
\end{figure}

Figure~\ref{fig1} illustrates the importance of accounting for subhalo infall 
over time. This figure is a scatter plot of mass versus pericentric distance
for two different satellite populations within all four galaxy-sized host CDM
halos. The masses and distances in Figure~\ref{fig1} have been scaled to 
the mass, $M_{\rm disk} = 3.53 \times 10^{10} \Mo$,
and radial scale length, $R_d = 2.82\kpc$, of the stellar disk in the primary  
galaxy model used in the satellite-disk encounter simulations. 
For the purposes of this presentation, we have also scaled 
the virial quantities ($M_{\rm vir}$ and $r_{\rm vir}$) of all four galaxy-sized 
host halos to a common host of {\it total} mass, $M_h=7.35 \times 10^{11} \Mo$, 
and {\it tidal} radius, $R_h=244.5 \kpc$, as in the primary disk galaxy model.
The dotted line encloses an area in the 
$M_{\rm sub}- r_{\rm peri}$ plane corresponding to subhalos more massive than $0.2 M_{\rm disk}$
with pericenters of $r_{\rm peri} \lesssim 20\kpc$ ($r_{\rm peri} \lesssim 7 R_d$). We refer to 
this area as the ``danger zone''.  Satellites within this area are expected to constitute effective 
perturbers and may cause considerable damage to the disk, but we intend this as a rough 
criterion to aid in illustrating our point.

The first satellite population in Figure~\ref{fig1} consists of the $z=0$ 
surviving substructures. The second subhalo population consists of systems 
that approach the central regions of their hosts since a redshift $z=1$.
These subhalos cross within a (scaled) infall radius of $r_{\rm inf}=50\kpc$ from the host halo center. 
This selection is fixed empirically to identify orbiting substructure that are likely to have 
a significant dynamical impact on the structure of the disk 
(\cite[Kazantzidis et al. 2007]{Kazantzidis_etal07}).
The masses associated with this group of satellites are defined at the simulation output 
time nearest the inward crossing of $r_{\rm inf}$. Pericenters are computed from 
the orbit of a test particle in a static NFW potential whose properties match 
those of the host CDM halo at the time of $r_{\rm inf}$.

Figure~\ref{fig1} demonstrates that the $z=0$ subhalo populations contain very few 
massive systems on potentially damaging orbits. In fact, statistics of all four galaxy-sized 
host halos indicate that only {\it one} satellite can be identified inside the danger zone
in this case. On the other hand, the danger zone contains numerous substructures that passed through or near 
the galactic disk since $z=1$. On average, $\sim 5$ satellites more massive 
than $0.2M_{\rm disk}$ cross through the central region of a galaxy-sized halo 
with $r_{\rm peri} \lesssim 20\kpc$ during this period. This suggests that close 
encounters between massive subhalos and galactic disks since $z=1$ are common 
occurrences in standard {\LCDM}. Thus, it is important to account for such accretion 
events to model the cumulative dynamical effects of halo substructure on disk galaxies.

In what follows, we focus on one of the host halo accretion histories to seed controlled 
$N$-body experiments of subhalo-disk encounters. We identify target satellites 
that are likely to have a substantial effect on the disk structure by imposing 
two selection criteria. First, we limit our search to satellites that approach
the central region of their host with small orbital pericenters 
($r_{\rm peri} \lesssim 20 \kpc$) since $z=1$. Second, we restrict
re-simulation to subhalos that are a significant fraction of the disk mass, 
but not more massive than the disk itself 
($0.2 M_{\rm disk} \lesssim M_{\rm sub} \lesssim M_{\rm disk}$).
The aforementioned criteria resulted in six accretion events of 
satellites with masses and tidal radii of 
$7.4 \times 10^{9} \lesssim M_{\rm sub}/M_{\odot} \lesssim 2 \times 10^{10}$,
and $r_{\rm tid} \gtrsim 20 \kpc$, respectively, from a single host to simulate 
over a $\sim 8$~Gyr period. Additional properties of these substructures can be 
found in \cite[Kazantzidis et al. (2007)]{Kazantzidis_etal07}. We modeled subhalo 
impacts onto the disk as a sequence of encounters. Starting with the first satellite, 
we included subsequent systems at the epoch when they were recorded in the cosmological 
simulation.

We extracted the density structures of these cosmological subhalos
and followed the procedure outlined in \cite[Kazantzidis et al. (2004)]
{Kazantzidis_etal04} to construct self-consistent, $N$-body
realizations of satellites models. Each system was represented with 
$N_{\rm sat} = 10^6$ particles and a gravitational softening length of
$\epsilon_{\rm sat}=150 \pc$. For the primary disk galaxy, we used $N_d=10^{6}$
particles to represent the disk, $N_b=5\times10^{5}$ in the bulge, 
and $N_h=2\times10^{6}$ in the dark matter halo, and softenings 
of $\epsilon_d=50 \pc$, $\epsilon_b=50 \pc$, and $\epsilon_h=100 \pc$, 
respectively. All satellite-disk encounter simulations were carried out 
using PKDGRAV (\cite[Stadel 2001]{Stadel01}). 

The ``final'' disk discussed in the next sections has experienced 
all six subhalo impacts and was evolved in isolation for $\sim 4$~Gyr 
after the last interaction.  This ensures that all of the
resultant morphological features are long-lived rather than transient.
Consequently, our results are relevant to systems that exhibit no obvious,
ongoing encounters. Finally, we compute all disk properties and show 
all visualizations of the disk morphology after centering the disk to its 
center of mass and rotating it to a new coordinate frame defined by the three 
principal axes of the total disk inertia tensor. 

\section{Global Disk Morphology}
\label{sub:morphology}

\begin{figure}
\begin{center}
 \includegraphics[width=4.5in]{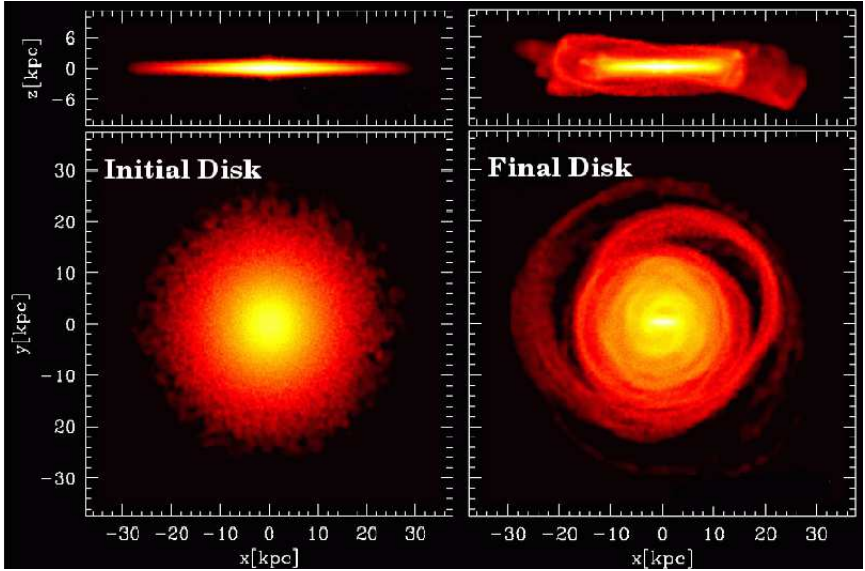} 
 \caption{Density maps of disk stars illustrating the global morphological transformation of a
   galactic disk subject to a {\LCDM}-motivated satellite accretion history. 
   The {\it left} panel shows the initial disk assuming that the sequence of satellite-disk 
   interactions initiates at $z=1$, while the {\it right} panel depicts the disk 
   after the last satellite passage, evolved in isolation for additional $\sim
   4$~Gyr, so that the evolution of disk stars is followed from $z=1$ to
   $z=0$. The edge-on ({\it upper panels}) and face-on ({\it bottom panels}) 
   views of the disk are displayed in each frame and the local density was
   calculated using an SPH smoothing kernel of $32$ particles. 
   Satellite-disk interactions of the kind expected in {\LCDM} models 
   produce several distinctive signatures in galactic disks including: 
   long-lived, low-surface brightness, ring-like features in the
   outskirts; conspicuous flares; bars; and faint filamentary structures
   above the disk plane that (spuriously) resemble tidal streams in configuration 
   space. These morphological features are similar to those being discovered in the Milky 
   Way, M31, and in other disk galaxies.}
   \label{fig2}
\end{center}
\end{figure}

Figure~\ref{fig2} depicts the transformation of the global structure of a thin 
galactic disk that experiences a merging history of the kind expected in the 
{\LCDM} paradigm of structure formation. 
This figure shows face-on and edge-on views of the initial and 
final distribution of disk stars. Particles are color-coded on 
a logarithmic scale with brighter colors in regions of higher 
stellar density.

Figure~\ref{fig2} demonstrates that encounters with CDM substructure are 
responsible for generating several distinctive morphological signatures in the disk.
The final disk is considerably thicker (or ``flared'') compared to the initial 
distribution of disk stars and a wealth of low-density features have developed 
both in and above the disk plane as a consequence of these disturbances.
Particularly intriguing is the fact that a high-density, in-plane structure survives after 
the satellite bombardment. A standard ``thin-thick'' disk decomposition analysis 
for the final disk indicates that this feature would be recognized 
as a thin disk component (\cite[Kazantzidis et al. 2007]{Kazantzidis_etal07}). 
The edge-on view of the final disk also reveals additional filamentary structures and other
complex configurations above the disk plane. These structures
bear some resemblance to tidal streams, but are in fact disk stars that have 
been excited by the subhalo impacts. Interestingly, the same
image shows a characteristic ``X'' shape in the bright
central disk, a finding also reported by \cite[Gauthier et
al. (2006)]{Gauthier_etal06}. This feature is often linked to secular evolution of galaxies
driven by the presence of a bar when it buckles as a result of becoming
unstable to bending modes.

The face-on image of the final disk illustrates the formation of a 
moderately strong bar and extended ring-like features in the outskirts 
of the disk. The existence of these features indicate 
that the axisymmetry of the disk has been destroyed by the encounters 
with the infalling subhalos. We emphasize that the
aforementioned structures are persistent, surviving for a
considerable time after the satellite passages ($\sim 4$~Gyr),
and that the bar is induced in response to the accretion events, 
not by amplified noise. Both features drive further evolution by redistributing 
mass and angular momentum throughout the disk.  Thus, encounters with infalling
satellites affect galactic disks both {\it directly} by imparting energy to stars 
and {\it indirectly} by exciting global instabilities. 
Lastly, the final face-on disk is significantly more structured at low 
densities and large radii compared to the initial disk and is quite
reminiscent of the outer disk structure seen around M31 
(\cite[Ibata et al. 2005]{Ibata_etal05}).

\section{Disk Thickening}
\label{sub:thickening}

\begin{figure}
\begin{center}
 \includegraphics[width=4in]{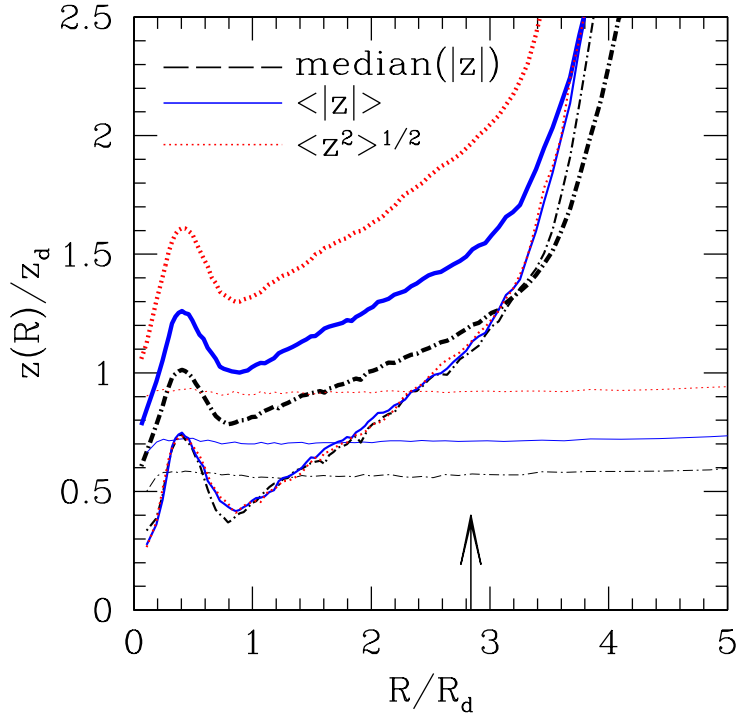} 
 \caption{Disk thickening. Thickness profiles, $z(R)$, for the disk initially ({\it thin lines})
  and after the satellite passages ({\it thick lines}).
  Lines of {\it intermediate} thickness show the fractional increase in disk thickness
  caused by the subhalo bombardment. Profiles are normalized to the initial
  disk scale height, $z_d$, and are plotted as a function of projected radius in
  units of the disk radial scale length, $R_d$. Different lines correspond to
  different measures of disk thickness. The arrow indicates the location of the solar radius, 
  $R_\odot$. The galactic disk thickens considerably at all radii as a result of the encounters 
  with CDM substructure.}
\label{fig3}
\end{center}
\end{figure}

Among the most striking signatures induced by the subhalo accretion events 
in our simulations is the pronounced increase in disk thickness. 
A quantitative analysis of disk thickening is presented in 
Figure~\ref{fig3}. This figure shows that the initial disk thickens 
considerably at all radii as a result of the substructure impacts.
Remarkably, the scale height of the disk near the solar radius increases 
in excess of a factor of $2$. The outer disk is much more susceptible to damage 
by the infalling satellites: at $R = R_d$ the scale height grows by $\sim 50\%$ 
compared to approximately a factor of $3$ increase at $R = 4 R_d$.
The larger binding energy of the inner, exponential disk and the presence of a massive 
bulge ($M_ b \sim 0.3 M_{\rm disk})$ that acts as a sink of satellite orbital energy are responsible for the robustness 
of the inner disk. Given that infalling subhalos are very extended and 
the self-gravity of the disk grows weaker as a function of distance from the center, 
it is not unexpected that the scale height of the disk should increase with radius yielding 
a flared disk. Indeed, making the simplest assumption that the accreting satellites deposit 
their orbital energy roughly constant in radius, it can be shown that the disk scale height 
will increase as $\Delta z(R) \propto \Sigma_d^{-2}(R)$, where $\Sigma_d(R)$ is the 
disk surface density (\cite[Kazantzidis et al. 2007]{Kazantzidis_etal07}).

\cite[Font et al. (2001)]{Font_etal01} and \cite[Gauthier et al. (2006)]{Gauthier_etal06} 
performed similar numerical studies of the dynamical response of disks to CDM subhalos.  
Both investigations reported negligible tidal effects on the global structure of the 
disk.  In contrast, Figure~\ref{fig3} indicates substantial disk thickening due to 
substructure bombardment.  
The primary reason for this discrepancy is that we followed 
the formation {\it history} of a host halo since $z \sim 1$, whereas 
\cite[Font et al. (2001)]{Font_etal01} and \cite[Gauthier et al. (2006)]{Gauthier_etal06} 
considered the $z=0$ population of surviving substructure present in a CDM
halo. As we mentioned above, this difference is critical because as subhalos on highly eccentric
orbits at early epochs continuously lose mass, the fraction of massive
satellites with small orbital pericenters that are most capable of severely
perturbing the disk declines with redshift so that few remain by $z=0$. 

%

\section{The Monoceros Ring}
\label{sub:monoceros}

The face-on view of the final disk in Figure~\ref{fig2} reveals the existence
of pronounced ring-like features at its outskirts. These 
low-density, long-lived dynamically-cold rings are nearly in-plane and are qualitatively
similar to the Monoceros ring, an intriguing stellar structure known to
extend around the MW (\cite[Newberg et al. 2002]{Newberg_etal02}). This coherent
feature spans $\sim 170^\circ$ degrees around the
Galaxy and lies near the disk plane at a Galacto-centric distance of
$\sim 20\kpc$ (e.g., \cite[Newberg et al. 2002]{Newberg_etal02}). 
It is unclear whether this structure is
yet another example of tidal debris from an accreted dwarf galaxy
(e.g., \cite[Yanny et al. 2003]{Yanny_etal03}) or a stellar extension of the disk itself
(e.g., \cite[Ibata et al. 2003]{Ibata_etal03}).

\begin{figure}
\begin{center}
  \includegraphics[height=2.4in]{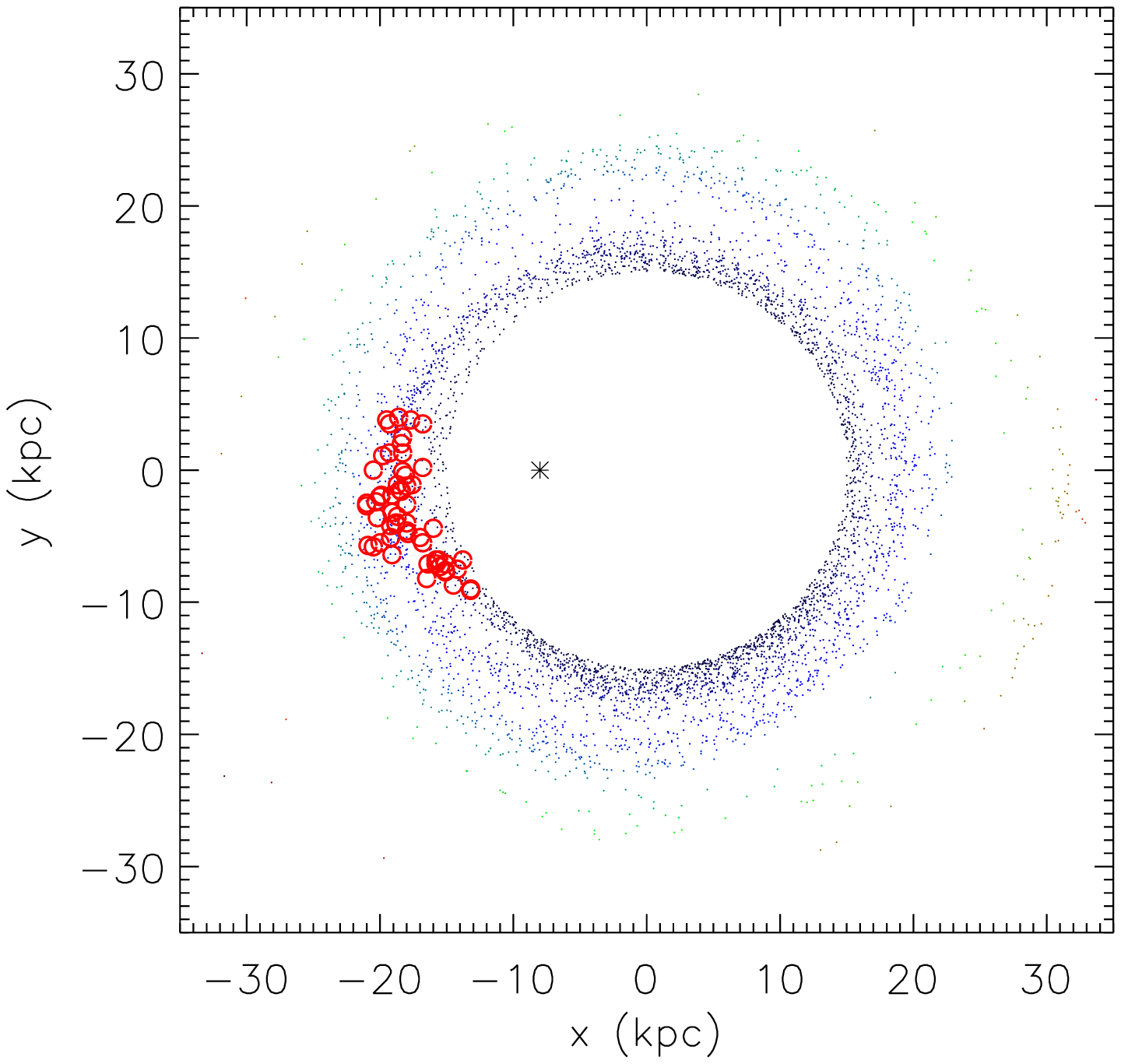}
   \includegraphics[height=2.4in]{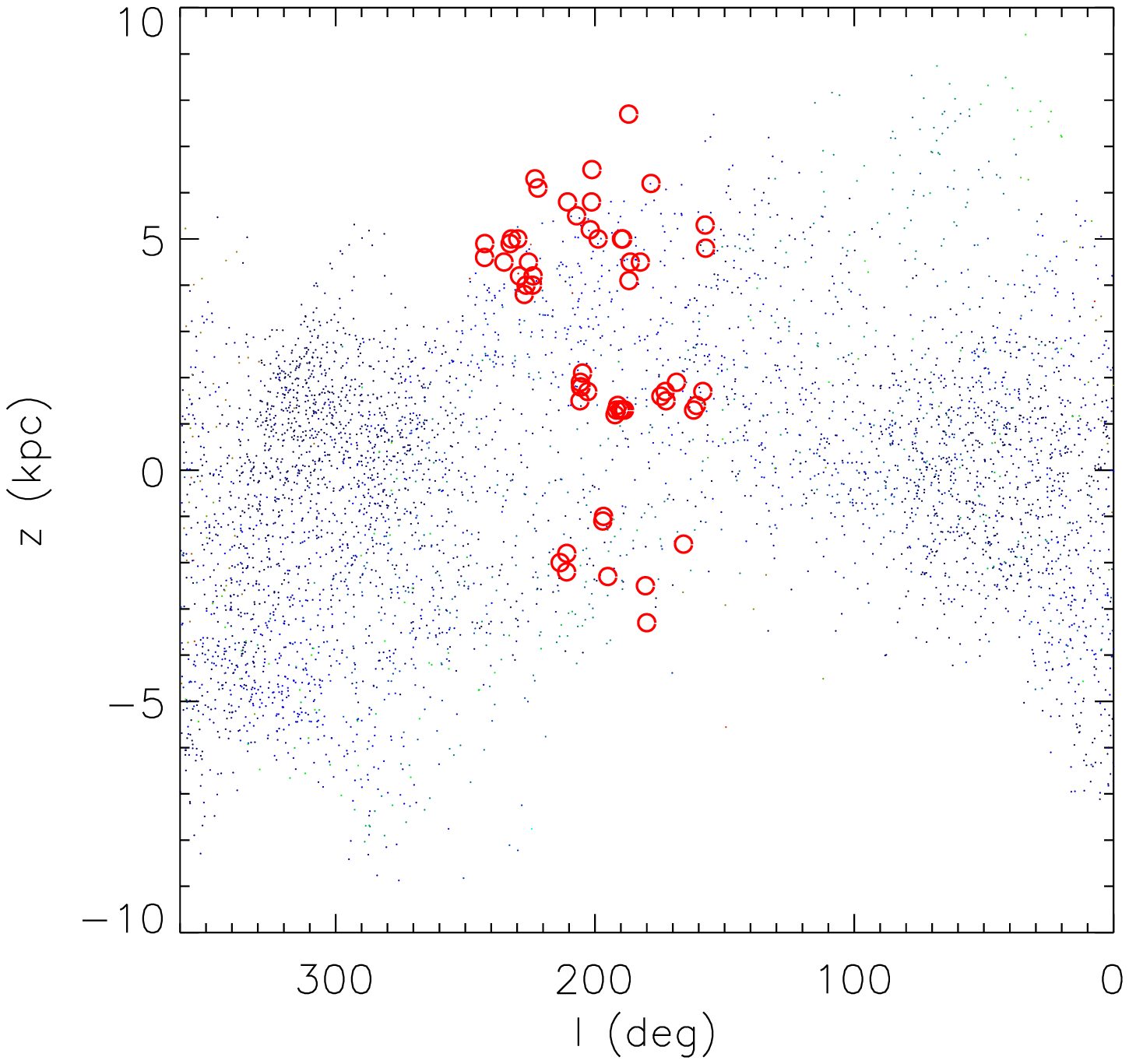} \\
    \includegraphics[height=2.4in]{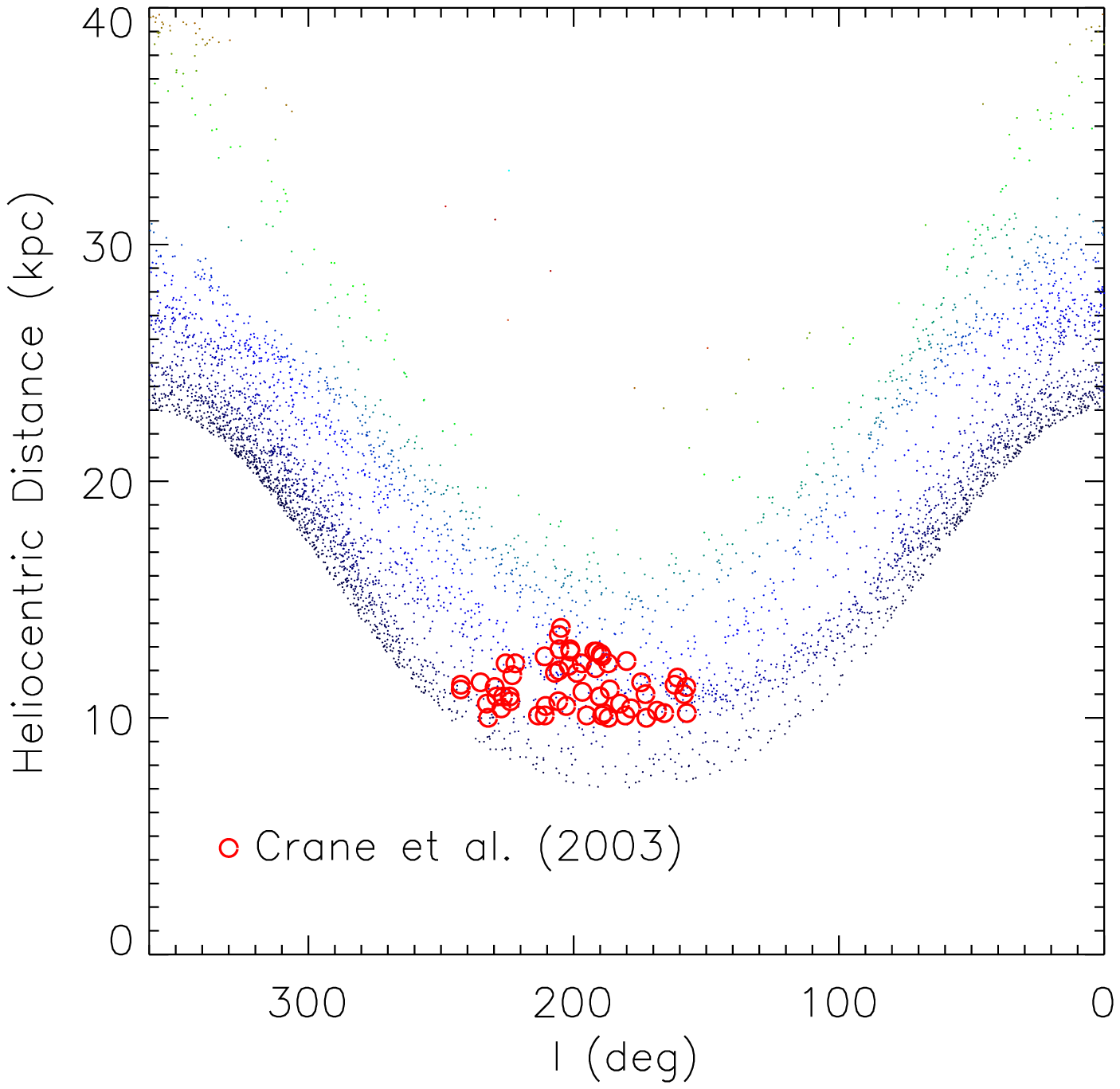}
     \includegraphics[height=2.4in]{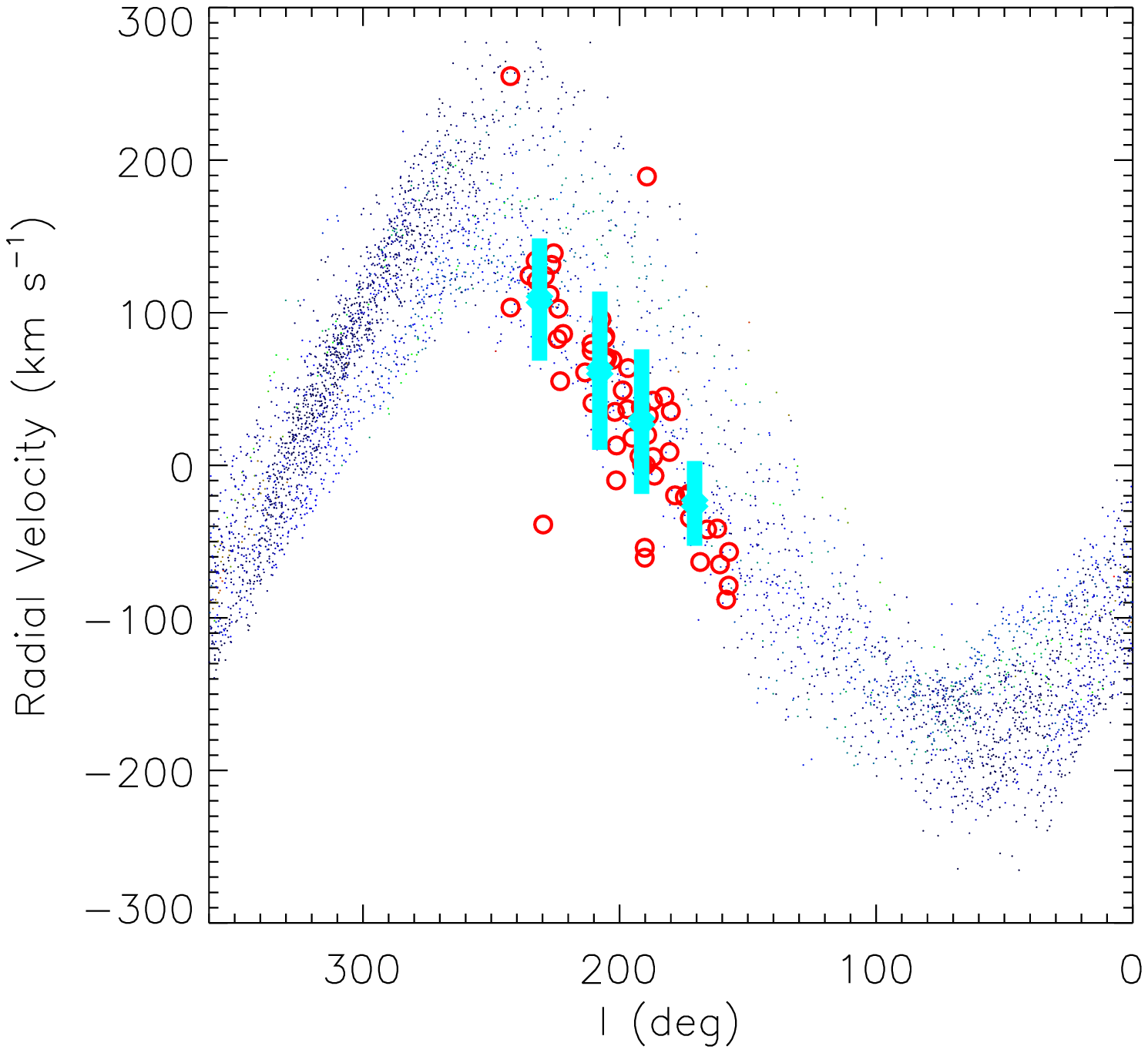}
\caption{Monoceros ring comparison. Circles are data of 
  M giant stars from the \cite[Crane et al. (2003)]{Crane_etal03} study of the Monoceros 
  stream in the direction of the Galactic anticenter. Points in each panel 
  correspond to stars in the outer regions ($R>15\kpc$) of the final 
  disk. These stars are color-coded according to their galacto-centric radius, 
  with dark blue/black at $R=15\kpc$, light-blue at $R\simeq 18\kpc$, 
  green at $R \simeq 22\kpc$, and orange/red at $R \gtrsim 25\kpc$.
  The {\it upper-left panel} shows a face-on view of the disk and the 
  asterisk marks the adopted position of the Sun at $(x,y,z) = (8, 0, 0)\kpc$.
  The {\it upper-right}, {\it lower-left}, and {\it lower-right} panels show vertical  
  distance of stars above the disk plane, $z$, heliocentric distance, and heliocentric
  radial velocity, respectively, as a function of Galactic longitude in degrees, $l$.   
  {\it Diamonds} in the lower-right panel correspond to the median line-of-sight velocities 
  in four bins in longitude $l = [240-210, 210-180, 180-150, 150-120]\degrees$
  for a specific stream-feature identified in the simulations. The associated bars
  reflect the line-of-sight velocity dispersion in each bin. Accretion
  histories of the kind expected in {\LCDM} models produce
  dynamically-cold, ring-like features around galactic disks that are
  quantitatively similar to the Monoceros ring in the MW.}
\label{fig4}
\end{center}
\end{figure}

The reminiscence of the resultant rings in our simulations to that of the 
Monoceros ring is suggestive that the latter 
may have been generated via an excitation of an ancient disk's stars. 
In order to check the general validity of this scenario, 
Figure~\ref{fig4} presents a direct comparison between the 
ring-like structures generated via satellite-disk interactions
and the Monoceros ring feature towards the Galactic anticenter. 
The circles in this figure correspond to a kinematic
study of M giant stars by \cite[Crane et al. (2003)]{Crane_etal03}, a follow-up to 
the \cite[Rocha-Pinto et al. (2003)]{Rocha-Pinto_etal03} 
effort to identify M giants associated with the ring. 
We choose the \cite[Crane et al. (2003)]{Crane_etal03}
sample because these data span the Monoceros stream uniformly 
over $\sim 100\degrees$ with both precise membership criteria and
good velocity determinations. For definiteness, we focus 
our analysis on the distribution of stars at galacto-centric radii larger
than $15\kpc$. These stars are color-coded according to their 
radial position from the center of the disk. Figure~\ref{fig4} shows vertical 
distance above the disk plane ($z$), heliocentric distance, and heliocentric radial 
velocity of these stars as a function of Galactic longitude, $l$.

It is worth emphasizing that the stars in our simulated data are more finely
sampled than those of \cite[Crane et al. (2003)]{Crane_etal03}. To facilitate 
further comparison, in the lower-right panel, we have more precisely  
identified a stream-feature along four bins in longitude $l = [240-210,  
210-180, 180-150, 150-120]\degrees$ and associated cuts in helio-centric distance 
$d_{\rm helio} = [14-12, 12.5-10, 12.5-10, 12.5-10]\kpc$.  
The four diamonds correspond to the median line-of-sight velocities and 
longitudes in each of these angular bins. The bars reflect the line-of-sight velocity 
dispersion in each bin, which is approximately $\sigma_{\rm los}\simeq 40\kms$ 
across the stream. We stress that using the \cite[Crane et al. (2003)]{Crane_etal03}
data across the first three of these bins we measured a line-of-sight velocity dispersion of
$\sigma_{\rm los} \simeq 39 \kms$, consistent with the corresponding values associated 
with the simulated data. 
Note that \cite[Crane et al. (2003)]{Crane_etal03} estimate a line-of-sight
velocity dispersion of $\sigma_{\rm los}= 20 \pm 4\kms$ for the ring, 
which differs significantly
from the aforementioned value of $\sigma_{\rm los} \simeq 39 \kms$.
This is because the former was calculated using a subset of $53$ of the
$58$ stars in their sample, obtained after
a $2.5\sigma$ ($\pm 50 \kms$) rejection threshold was applied
about a third-order polynomial fit to the velocity trend. When we perform
a similar rejection threshold within each of our four angular
bins we find $\sigma_{\rm los} \sim 24.5\kms$ for our simulated ring, a 
value consistent with the quoted velocity dispersion for the Monoceros stream 
from \cite[Crane et al. (2003)]{Crane_etal03}.

The general agreement in the properties of rings 
generated in our simulations with those of the Monoceros ring structure is 
encouraging. This agreement is particularly noteworthy because our simulation program 
did {\it not} aim to reproduce such a feature. Our simulation campaign has explored only one realization of 
a galaxy-sized halo accretion history and we tuned no aspect of the computations to produce 
this agreement. This is a significant point that bears emphasis.  
One may contrast our model of a disk origin for the 
Monoceros stream to previous models that have attempted to explain this structure via the 
accretion and disruption of a satellite galaxy
(\cite[Pe{\~n}arrubia et al. 2005]{Penarrubia_etal05}). Neither our model nor the model of 
\cite[Pe{\~n}arrubia et al. (2005)]{Penarrubia_etal05} can be falsified by extant data; however, our ring was produced 
as an unforeseen byproduct of our simulation campaign while the 
\cite[Pe{\~n}arrubia et al. (2005)]{Penarrubia_etal05} ring was produced as part of 
a program aimed at tuning a satellite 
accretion event to yield a structure similar to the Monoceros ring. Lastly,
it is important to emphasize that accreted-star models seem to
require accretion events on nearly-circular, low-inclination orbits. Orbits of this
type are quite rare for satellites that penetrate deep into the host halo 
(e.g., \cite[Ghigna et al. 1998]{Ghigna_etal98}). Our disk-excitation scenario 
thus constitutes a viable alternative for the origin of the Monoceros ring in the 
context of {\LCDM}.

It is crucial to be able to distinguish between these alternative mechanisms for 
generating structures such as the Monoceros ring. In principle, the metallicity distribution of stars in the stream 
can be used to constrain the competing models. The stars in an accreted
satellite would generally be expected to have different metallicity from disk 
stars, while in our model the metallicity of the stream 
should be comparable to the metallicity of the thick disk stars 
surrounding the stream. Unfortunately, metallicity estimates of Monoceros ring stars span a range 
of values and there is no definitive conclusion that can be reached at 
present. Similar uncertainties exist for the characterization of the 
outer thick disk. If the outer thick disk is as metal
rich as the canonical thick-disk value, $[{\rm Fe/H}] \approx -0.7$ to $-1$ 
(\cite[Gilmore et al. 1995; Ivezi{\'c} et al. 2008]{Gilmore_etal95,Ivezic_etal08}),
the higher ring metallicities estimated by \cite[Crane et
al. (2003)]{Crane_etal03} and \cite[Ivezi{\'c} et al. (2008)]{Ivezic_etal08}
would be consistent with our scenario, while the lower ring metallicities 
reported by \cite[Yanny et al. (2003)]{Yanny_etal03} would be discordant.  
Overall, the robust utilization of metallicity constraints requires 
refining the observational measurements for the Monoceros ring and 
securing the metallicity spread in the outer thick disk. Moreover, precise 
predictions would require simulations of star formation and metal enrichment.

Another promising method that may falsify a disk-excitation scenario 
is to identify ring populations that are not expected to pre-exist in the outer 
disk. Interestingly, there have been indications that some globular clusters may be associated with the 
Monoceros stream (\cite[Crane et al. 2003]{Crane_etal03}). While this association
is still a matter of debate, the existence of such globular clusters 
would point to an external origin for the Monoceros structure. Nevertheless, 
the dynamical study we present is robust and though disk-excitation by substructure 
may not be the source of the Monoceros feature, low-surface brightness features of 
this kind should not be uncommon about disk galaxies.


\begin{thebibliography}{}

\bibitem[{Blumenthal} {et~al.}(1984)]{Blumenthal_etal84}
{Blumenthal}, G.~R., {Faber}, S.~M., {Primack}, J.~R., \& {Rees}, M.~J. 1984,
\textit{Nature}, 311, 517

\bibitem[{{Crane} {et~al.}(2003){Crane}, {Majewski}, {Rocha-Pinto},
  {Frinchaboy}, {Skrutskie}, \& {Law}}]{Crane_etal03}
{Crane}, J.~D., {Majewski}, S.~R., {Rocha-Pinto}, H.~J., {Frinchaboy}, P.~M.,
  {Skrutskie}, M.~F., \& {Law}, D.~R. 2003, \textit{Ap. Lett.}, 594, L119

\bibitem[{{Font} {et~al.}(2001){Font}, {Navarro}, {Stadel}, \&
  {Quinn}}]{Font_etal01}
{Font}, A.~S., {Navarro}, J.~F., {Stadel}, J., \& {Quinn}, T. 2001, \textit{Ap. Lett.}, 563,
  L1

\bibitem[{{Gauthier} {et~al.}(2006){Gauthier}, {Dubinski}, \&
  {Widrow}}]{Gauthier_etal06}
{Gauthier}, J.-R., {Dubinski}, J., \& {Widrow}, L.~M. 2006, \textit{ApJ}, 653, 1180

\bibitem[{{Ghigna} {et~al.}(1998){Ghigna}, {Moore}, {Governato}, {Lake},
  {Quinn}, \& {Stadel}}]{Ghigna_etal98}
{Ghigna}, S., {Moore}, B., {Governato}, F., {Lake}, G., {Quinn}, T., \&
  {Stadel}, J. 1998, \textit{MNRAS}, 300, 146

\bibitem[{{Gilmore} {et~al.}(1995){Gilmore}, {Wyse}, \&
  {Jones}}]{Gilmore_etal95}
{Gilmore}, G., {Wyse}, R.~F.~G., \& {Jones}, J.~B. 1995, \textit{AJ}, 109, 1095

\bibitem[{{Ibata} {et~al.}(2005){Ibata}, {Chapman}, {Ferguson}, {Lewis},
  {Irwin}, \& {Tanvir}}]{Ibata_etal05}
{Ibata}, R., {Chapman}, S., {Ferguson}, A.~M.~N., {Lewis}, G., {Irwin}, M., \&
  {Tanvir}, N. 2005, \textit{ApJ}, 634, 287

\bibitem[{{Ibata} {et~al.}(2003){Ibata}, {Irwin}, {Lewis}, {Ferguson}, \&
  {Tanvir}}]{Ibata_etal03}
{Ibata}, R.~A., {Irwin}, M.~J., {Lewis}, G.~F., {Ferguson}, A.~M.~N., \&
  {Tanvir}, N. 2003, \textit{MNRAS}, 340, L21

\bibitem[{{Ivezi{\'c}} {et~al.}(2008)}]{Ivezic_etal08}
 {Ivezi{\'c}}, {\v Z}. {et~al.} 2008, \textit{ApJ} accepted (astro-ph/0804.3850)

\bibitem[{{Kazantzidis} {et~al.}(2007){Kazantzidis}, {Bullock}, {Zentner},
  {Kravtsov}, \& {Moustakas}}]{Kazantzidis_etal07}
{Kazantzidis}, S., {Bullock}, J.~S., {Zentner}, A.~R., {Kravtsov}, A.~V., \&
  {Moustakas}, L.~A. 2007, \textit{ApJ} accepted (astro-ph/0708.1949)

\bibitem[{{Kazantzidis} {et~al.}(2004){Kazantzidis}, {Magorrian},
  \& {Moore}}]{Kazantzidis_etal04}
{Kazantzidis}, S., {Magorrian}, J., \& {Moore}, B. 2004, \textit{ApJ}, 601, 37

\bibitem[{{Klypin} {et~al.}(1999){Klypin}, {Kravtsov}, {Valenzuela}, \&
  {Prada}}]{Klypin_etal99}
{Klypin}, A., {Kravtsov}, A.~V., {Valenzuela}, O., \& {Prada}, F. 1999, \textit{ApJ},
  522, 82

\bibitem[{{Kravtsov}(1999)}]{Kravtsov99}
{Kravtsov}, A.~V. 1999, PhD thesis, New Mexico State University

\bibitem[{{Newberg} {et~al.}(2002)}]{Newberg_etal02}
{Newberg}, H.~J. {et~al.} 2002, \textit{ApJ}, 569, 245

\bibitem[{{Pe{\~n}arrubia} {et~al.}(2005)}]{Penarrubia_etal05}
{Pe{\~n}arrubia}, J. {et~al.} 2005, \textit{ApJ}, 626, 128

\bibitem[{{Quinn} \& {Goodman}(1986)}]{Quinn_Goodman86}
{Quinn}, P.~J. \& {Goodman}, J. 1986, \textit{ApJ}, 309, 472

\bibitem[{{Read} {et~al.}(2008){Read}, {Lake}, {Agertz}, \&
  {Debattista}}]{Read_etal08}
{Read}, J.~I., {Lake}, G., {Agertz}, O., \& {Debattista}, V.~P. 2008, \textit{MNRAS}
  accepted (astro-ph/0803.2714)

\bibitem[{{Rocha-Pinto} {et~al.}(2003){Rocha-Pinto}, {Majewski}, {Skrutskie},
  \& {Crane}}]{Rocha-Pinto_etal03}
{Rocha-Pinto}, H.~J., {Majewski}, S.~R., {Skrutskie}, M.~F., \& {Crane}, J.~D.
  2003, \textit{Ap. Lett.}, 594, L115

\bibitem[{{Stadel}(2001)}]{Stadel01}
{Stadel}, J.~G. 2001, Ph.D.~Thesis, Univ. of Washington

\bibitem[{{Velazquez} \& {White}(1999)}]{Velazquez_White99}
{Velazquez}, H. \& {White}, S.~D.~M. 1999, \textit{MNRAS}, 304, 254

\bibitem[{{Villalobos} \& {Helmi}(2008)}]{Villalobos_Helmi08}
{Villalobos}, {\'A}. \& {Helmi}, A. 2008, \textit{MNRAS} submitted (astro-ph/0803.2323)

\bibitem[{{Widrow} \& {Dubinski}(2005)}]{Widrow_Dubinski05}
{Widrow}, L.~M. \& {Dubinski}, J. 2005, \textit{ApJ}, 631, 838

\bibitem[{{Yanny} {et~al.}(2003)}]{Yanny_etal03}
{Yanny}, B. {et~al.} 2003, \textit{ApJ}, 588, 824

\bibitem[{{Zentner} \& {Bullock}(2003)}]{Zentner_Bullock03}
{Zentner}, A.~R. \& {Bullock}, J.~S. 2003, \textit{ApJ}, 598, 49


\end{thebibliography}
\end{document}